\documentclass[fleqn,twoside]{article}
\usepackage{espcrc2}

\usepackage{graphicx}

\usepackage{epsfig}
\usepackage{amsfonts}
\usepackage[figuresright]{rotating}
\newcommand{\AmS}{{\protect\the\textfont2
  A\kern-.1667em\lower.5ex\hbox{M}\kern-.125emS}}

\def\half{{\scriptstyle \raise.2ex\hbox{${1\over2}$}}}
\def\fourth{{\scriptstyle \raise.2ex\hbox{${1\over4}$}}}

\newcommand*{\chpt}{\raise0.4ex\hbox{$\chi$}PT}
\newcommand*{\schpt}{S\raise0.4ex\hbox{$\chi$}PT}
\newcommand*{\ie}{\textit{i.e.},\ }
\newcommand*{\eg}{\textit{e.g.},\ }
\newcommand*{\etc}{\textit{etc.}}

\newcommand*{\et}{\textit{et al.}}

\newcommand*{\npb}[1]{Nucl.\ Phys.\ \textbf{B#1}}
\newcommand*{\prd}[1]{Phys.\ Rev.\ \textbf{D#1}}

\newcommand*{\boston}{Nucl.\ Phys.\ {\bf B}
 (Proc.\ Suppl.) {\bf 119} (2003)}

\newcommand*{\fm}{{\rm fm}}

\newcommand*{\Tr}{\textrm{Tr}}
\newcommand*{\tr}{\textrm{tr}}

\newcommand{\cD}{\mathcal{D}}

\newcommand{\cL}{\mathcal{L}}
\newcommand{\cM}{\mathcal{M}}

\newcommand{\cO}{\mathcal{O}}

\newcommand{\cU}{\mathcal{U}}
\newcommand{\cV}{\mathcal{V}}

\def\eqs#1#2{Eqs.~(\ref{eq:#1}) and (\ref{eq:#2})}

\title{Staggered Chiral Perturbation Theory
  \thanks{talk presented by C.\ Aubin at \textit{Lattice 2003};
     to be published in the proceedings}}
\author{C. Aubin$^{\rm a}$, C. Bernard
\address{Department of Physics, 
Washington University, St. Louis, MO 63130, USA}}

\begin{document}

\begin{abstract}
We discuss how to formulate a staggered chiral perturbation
 theory (\schpt). This
amounts to a generalization of the Lee-Sharpe Lagrangian to include more
than one flavor (\ie\ multiple staggered fields), which turns out to be
nontrivial. 
One loop
corrections to pion and kaon masses and decay constants are computed as
examples in three cases: the quenched, partially quenched, and full
(unquenched) case. 
The results for the one loop mass and decay
constant corrections have
already been presented in 
Ref.~\cite{SCHPT}.
\vspace{-1.5pc}
\end{abstract}

\maketitle

In order to reproduce the chiral behavior of staggered fermion
simulation data as $a^2\to 0$, one must account for the systematic
effects arising from $\cO(a^2)$ taste violations. 
(\textit{Taste} refers
to the staggered degrees of freedom resulting from doubling, while
\textit{flavor} refers to true quark flavor.) These taste violations
are not negligible at current lattice spacings ($a\approx 0.9-0.12\ \fm$
\cite{MILC-RECENT}).
Lee and Sharpe have formulated \cite{LEE-SHARPE} such a staggered
chiral perturbation theory (\schpt) for one staggered field.  Here we
describe the generalization to the case of $n$ flavors.

We follow a three-step procedure:
{\small
\begin{itemize}
\item First, we generalize the Lee-Sharpe
  Lagrangian to incorporate multiple flavors. This is the
  ``$4\!+\!4\!+\!...$'' theory---$n$ flavors with 4 tastes per flavor.
\item Next we calculate various meson properties; specifically we will
  calculate the Goldstone pion mass and decay constant for $n=3$.
\item Finally, we adjust this result by hand to keep only one taste
  per flavor,
  thus accounting for
  taking the $\root 4 \of {\rm Det}$ in simulations \cite{MILC-RECENT}.
\end{itemize} }
This can be done for any of the following cases: partially quenched
($m_{\rm valence} \ne m_{\rm sea}$), ``full QCD'' ($m_{\rm valence} =
m_{\rm sea}$) and quenched ($m_{\rm valence} \ne m_{\rm sea}$,
$m_{sea}\to\infty$); here we will show results for the partially
quenched case.  Complete results have been reported in
Ref.~\cite{SCHPT} and preliminary fits to simulation data are shown in
Ref.~\cite{CB-POSTER}.

We begin here by formulating \schpt\ for $n$ flavors, stating only the
differences between the $n=1$ \cite{LEE-SHARPE}
and $n>1$ cases. We collect the Goldstone
bosons arising from the spontaneous breakdown of $SU(4n)_L\times
SU(4n)_R \to SU(4n)_{\rm vec}$ in the unitary field $\Sigma =
\exp(i\Phi / f)$, with $\Phi$ the hermitian $4n\times 4n$ matrix (shown
is the $n=3$ case):
\vspace{-.5pc}
\begin{eqnarray}
      \Phi = \left( \begin{array}{ccc}
     	U  & \pi^+ & K^+ \\*
     	\pi^- & D & K^0  \\*
	K^-  & \bar{K^0}  & S  \end{array} \right) ,
\end{eqnarray}
where $U = U_a T_a$, $K^+ = K^+_a T_a$, \etc\ are each $4\times 4$
matrices, and
the $T_a$ are the 16 taste matrices $\xi_5$, $\xi_\mu$,
\etc\ We include a $4n\times 4n$ mass matrix 
$\cM = \textrm{diag}(m_u I, m_d I, m_s I, ...)$, with $I$ the
$4\times 4$ identity matrix.

We have the Lagrangian
\begin{eqnarray}
    \cL &\!\! = \!\!& \frac{f^2}{8} \Tr(\partial_{\mu}\Sigma
    \partial_{\mu}\Sigma^{\dagger}) - \frac{1}{4}\mu f^2
    \Tr(\cM\Sigma + \cM\Sigma^{\dagger}) \nonumber\\*
    &&+ \cL_{\rm singlet} +
    a^2\cV ,
\end{eqnarray}
where the first two terms are the standard kinetic and mass terms from
continuum \chpt\ \cite{GASSER-LEUTWYLER}.
$\cL_{\rm singlet}$, proportional to $m^2_0$, arises from
the anomaly, and gives the $SU(4n)$ singlets a large mass
(we will take $m_0\to\infty$ in the end).
The last term is the
taste-symmetry breaking potential; for a single flavor, Lee and Sharpe
found this to be the sum of six operators: 
\begin{eqnarray}
    -\cV \!\!\! & = &\!\!\! \sum_{i=1}^6 C_i O_i = C_1
    \tr(\xi_5\Sigma\xi_5\Sigma^{\dagger}) +
    C_2\frac{1}{2} [ \tr(\Sigma^2) \nonumber\\*&&- 
      \tr(\xi_5\Sigma\xi_5\Sigma) + h.c.] + ... \ .
\end{eqnarray} 
The $O_i$ are taste-breaking chiral operators that correspond to
various four-quark operators. To get all terms in the potential
in single-trace form, Lee and Sharpe performed a Fierz
transformation on the original operators.

Because of this Fierz transformation, generalizing to multiple flavors
is tricky. Returning to the quark level, we recall that these
operators come from four-quark operators with a net momentum change of
$\cO(\pi)$, which changes quark taste. 
This gluon exchange can also
change color, but not flavor, so all four-quark operators must be of
the flavor-unmixed form: $\bar{q}_i (\gamma_s\otimes\xi_t)
q_i\bar{q}_j (\gamma_{s'}\otimes\xi_{t'}) q_j$, where
$(\gamma_s\otimes\xi_t)$ is the (spin$\otimes$taste)
matrix. In the naive theory, each bilinear is separately chirally
invariant---these are the ``odd'' bilinears in the staggered
theory. Thus, only the odd-odd four-quark operators are relevant here.

Keeping all of the four-quark operators in the flavor-unmixed form, we
then see that, using the standard spurion analysis,
the taste
matrices $\xi_t$ are singlets under the flavor $SU(n)$ symmetry 
\cite{CB-SChPT1}. This
means we can make the replacement $\xi_t \to
\left(\xi_t^{(n)}\right)_{ij} = \xi_t \delta_{ij}$, where $i$ and $j$
are flavor indices. This must be done before the Fierz transformation
performed by Lee and Sharpe to put the operators into single-trace
form. Only then do the chiral operators follow from the flavor-unmixed
four-quark operators.

For the operators $O_1,O_3,O_4$ and $O_6$ (which we combine into ``$\cU$''),
we can just replace $\xi_t\to\xi^{(n)}_t$. Instead of the operators
$O_2$ and $O_5$ we have four operators,
$O_{2V},O_{2A},O_{5V}$ and $O_{5A}$ (
``$\cU\,'$''),
which are not in single-trace form. For example: 
$O_{2V} = \frac{1}{4}[ \Tr(\xi^{(n)}_{\nu}\Sigma)
    \Tr(\xi^{(n)}_{\nu}\Sigma)  + h.c.]$ and 
$O_{2A} = \frac{1}{4} [ \Tr(\xi^{(n)}_{\nu
      5}\Sigma)\Tr(\xi^{(n)}_{5\nu}\Sigma)  + h.c.] $.
The full potential is then $\cV = \cU + \cU\,'$.

One of the consequences of the two-trace form of the terms in $\cU\,'$
is the appearance of quark-level hairpin terms similar to $\cL_{\rm
singlet}$.
These terms are of the form $+{\lambda_t'\over2}
(U_t+D_t+S_t+\cdots)^2$, where $\lambda'_t$ depends on the taste channel
we're discussing ($t=V$, $A$, or $I$): 
$\lambda'_I = 4m_0^2/3$ for the singlet case, and
for the vector and axial-vector cases $\lambda'_{V(A)} =
a^2\delta'_{V(A)}$, with $\delta'_{V(A)}$ a linear combination of
the coefficients in $\cU\,'$ \cite{SCHPT}.

\begin{figure}
    \includegraphics[width=2.8in,height=.5in]{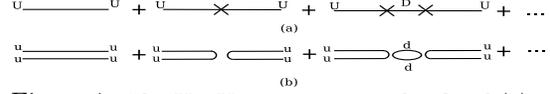}
    \vspace{-2.7pc}\caption{\small The $U-U$ propagator at the chiral (a) and
      quark (b) level. Each
      $\times$ corresponds to an insertion of $\lambda'_t$, and the
      intermediate meson could be $U$, $D$ or $S$ (for $n=3$).}
    \label{fig:hairpin}
  \vspace{-1.5pc}
\end{figure}

These hairpins allow mixing among the flavor-neutral mesons
(shown in Fig.~\ref{fig:hairpin}), which we can resum
\cite{SHARPE-SHORESH} to give a 
non-diagonal propagator:
\vspace{-.2pc}
\begin{eqnarray}
  G_{MN}
 &= & \frac{\delta_{MN}}{q^2 + m_{M}^2} -
 \frac{\lambda'}{(q^2 + m_{M}^2)(q^2 + m_{N}^2)}\nonumber\\ &&
 \times\left(\frac{(q^2 + m_{U}^2)(q^2 + m_{D}^2)\cdots}
      {(q^2 + m_{\pi^0}^2)(q^2 + m_{\eta}^2)\cdots}\right)\ ,
\end{eqnarray}
where the second term we denote by $\cD_{MN}$.

It is interesting to note that the symmetries of the multiple-flavor theory
are only slightly modified from those of the single-flavor theory. For
example:
{\small 
\begin{itemize}
\item The spontaneously broken symmetry ($m=0$ and $a=0$): $SU(4n)_L
  \times SU(4n)_R \to SU(4n)_{\rm vec}$, for any $n\ge 1$.
\item The residual chiral symmetry ($m=0$ and $a\ne 0$): 
  $U(n)_\ell\times U(n)_r$. 
  We use $\ell$ and $r$ (not $L$ and $R$) to denote the left and right
  symmetries to remind us that they are mixtures of chiral spin and
  taste.
\item Fermion number [$m\ne 0$ (for $n>1$, we assume all $n$ masses
  are nondegenerate) and $a\ne 0$]: $U(1)_{\rm  VEC}$ for one flavor,
  and this becomes $(U(1)_{\rm vec})^n$ for more than one flavor.
\end{itemize}}

We show here
the $n=3$ partially quenched result for the
pion mass and decay constant. Here a ``pion'' is any 
flavor-charged meson, and $n$ refers to the number of sea quarks.
We add two quenched
valence quarks, $x$ and $y$, and calculate
the properties of the Goldstone ``pion'' $P^+_5 = x\bar{y}$. $X$ and
$Y$ will refer to the flavor-neutral mesons composed of these quarks.

For the pion mass, we calculate the pion self energy evaluated at $p^2
= -m^2_{P^+_5}$. All connected (at the quark level) terms cancel and
we are left only with terms involving the disconnected terms.
To one loop, we have:
\begin{eqnarray}\label{eq:m2pi}
  \frac{m^2_{P^+_5}}{(m_x + m_y)}&\!\! =\!\! &
  \mu\biggl\{ 1+ \frac{1}{16\pi^2 f^2} \int
  \frac{d^4 q}{\pi^2} \biggl(2 \cD^V_{XY}+\nonumber\\&&
  2 \cD^{A}_{XY} 
  -\frac{1}{2}\cD^I_{XY}\biggr) + \textrm{[a.t.]}\biggr\} .
\end{eqnarray}
In this expression and below we leave off the analytic terms (denoted
by [a.t.]) for simplicity. Note also that
the pion mass vanishes in the chiral limit ($m_x$, $m_y\to 0$) as it
must, since this is the true Goldstone boson.

For the pion decay constant, we calculate the matrix element $\langle 0 |
j_{\mu5} | P^+_5(p) \rangle =-i f_{P^+_5}
p_{\mu}$, where $j_{\mu5}$ is the axial current for $P^+_5$. To one
loop we find
\begin{eqnarray}
  f_{P^+_5} & = & f\left(1 + \frac{1}{16\pi^2f^2}\delta\! f_{P^+_5}
  + \textrm{[a.t.]}\right)\ ,\\
  \delta\! f_{P^+_5} & = &
  -\frac{1}{8}\int\frac{d^4 q}{\pi^2} \Biggl[    
    \sum_{Q,t} \biggl( \frac{1}{q^2 + m^2_{Q_t}}\biggr)\nonumber\\
    &&+ \cD^I_{XX} - 2\cD^I_{XY} +
    \cD^I_{YY}+  \Bigl(4\cD^V_{XX} \nonumber\\
    && +8\cD^V_{XY}
    + 4\cD^{V}_{YY}\Bigr)+  \left( V\to A\right)
    \Biggr]\ .\label{eq:fpi}
\end{eqnarray}
The sum over $Q$ is over mesons with one sea and one valence quark,
and $t$ runs over the 16 tastes. 

These expressions are quite general, and any relevant result
can be found by taking limits before performing the
momentum integrals. For the ``full QCD'' 
case, set $m_x=m_u$ and $m_y=m_d$ ($m_s$) for
the true pion (kaon); for the
quenched case, take $m_{u,d,s}\to\infty$.

Once one takes the desired limits, one must perform the integrals in
\eqs{m2pi}{fpi}. The integrands are ratios of products of terms of the form
$(q^2+m^2)$. These can be expanded as sums of poles times their
residues. Performing the integrals, we keep only the chiral logarithms
(the analytic pieces are absorbed into ``[a.t.]'', and
we leave off finite volume effects here).
For a single pole, we get a term 
$\ell( m^2) \equiv  m^2 \ln (m^2/\Lambda^2)
$,
while for a double pole, we get 
$\tilde \ell(m^2) \equiv  -\left(\ln (m^2/\Lambda^2) + 1\right)
$ (with $\Lambda$ the chiral scale).
The residues multiplying these logarithms are complicated, and
are given in full detail along with the analytic terms
in Ref.~\cite{SCHPT}. Also, see Ref.~\cite{CB-POSTER} for
explicit expressions for $m_K^2$
and $f_\pi$.

The last step before having the final result for the mass and decay
constant is to adjust from four to one tastes per flavor. This is 
done using the ``quark flow technique,'' in Ref.~\cite{SCHPT}, where 
we determine where quark loops
arise, and multiply each corresponding loop by
$1/4$. However, from that approach it is not clear that this works
 at all orders in
perturbation theory. Further, even if it does work, the application of the
quark flow technique could be quite complex.

One can use the replica method
to automate this: Take $n_i$ the
quarks of each flavor $i$.
Then calculate quantities of interest to a given number of loops
 as analytic
functions of the $n_i$; in the end, set each $n_i = 1/4$. This should
take into account the transition from $4\to 1$ tastes per flavor at
all orders automatically.

Another interesting possibility that may arise in \schpt\ is that of
an unusual phase. If (with $a^2\Delta_A = m^2_{S_A} - m^2_{S_5}$)
\begin{equation}
  \delta'_A < \delta'_{A,\,{\rm crit}} 
  \equiv  -4\Delta_A\; \frac{1 + a^2\Delta_A/m^2_{S_5}}
	  {2 + 3a^2\Delta_A/m^2_{S_5}} ,
\end{equation}
$m^2_{\eta_A}$ could become negative for small, but non-zero
$m_u=m_d$. From fits, this does not appear likely for the physical
case of QCD. A corresponding condition for $m_u=m_d=m_s$ may be
satisfied, although such a phase would disappear in the continuum
limit. The possibility of an unusual phase 
requires further study \cite{AUBIN-CB}.

Calculations for 
$m_{P^+_5}^2$ and $f_{P^+_5}$ 
are complete for the
partially quenched, full and quenched cases. 
Preliminary fits are shown in Ref.~\cite{CB-POSTER},
and appear promising. The next
goal is to include heavy quarks so as to calculate
heavy-light decay constants, and an extension to baryons \cite{AUBIN-CB}.

This work was supported by the U.S.\ DOE.

\end{document}